# Two options for muon–proton collider at FNAL

Burak Dagli[1*], Umit Kaya[2], Arif Ozturk[1,2], Saleh Sultansoy[1,3]

[1]TOBB University of Economics and Technology, Ankara, Türkiye

[2]Ankara University, Institute of Accelerator Technologies, Ankara, Türkiye

[3]ANAS Institute of Physics, Baku, Azerbaijan

## Abstract

We discuss a possibility to construct multi-TeV scale $\mu p$ collider at FNAL. Main advantage of this project is existence of two ring tunnels tangential to each other. There are two possible options, namely, muons in main injector with protons in Tevatron ring and vice versa. Two choices are considered for center-of-mass energy values: 2.57 TeV using 8 T bending magnets and 5.10 TeV with 16 T magnets. It is shown that luminosity values exceeding $10^{33}$ cm$^{-2}$s$^{-1}$ can be achieved. It is obvious that µ-FNAL will provide unique potential for both SM (especially QCD basics) and BSM searches, far beyond the eRHIC and LHeC capacities.

**Contents**

*Corresponding Author: burakdagli@etu.edu.tr

# 1. Introduction

Lepton-hadron collisions play a crucial role in our understanding of matter's structure: proton form-factors were first observed in electron scattering experiments [1, 2], quarks were first observed at SLAC deep inelastic electron scattering experiments [3, 4], EMC effect was observed at CERN in deep inelastic muon scattering experiments [5] and so on. HERA, the first electron-proton collider, further explored structure of protons and provided parton distribution functions (PDFs) for the LHC and Tevatron.

Undoubtedly, the construction of TeV (or even multi-TeV) scale lepton-hadron colliders will be crucial to elucidate the fundamentals of QCD, which is responsible for 98% of the mass of the visible part of our Universe. It is worth to note that the electro-weak part of the Standard Model (SM) has been completed with discovery of Higgs boson at the LHC, but this is not the case for QCD part: the confinement hypothesis should be clarified. It should be emphasized that future lepton-hadron colliders will give opportunity to shed light on quark → hadron → nucleus transitions as well. Besides, construction of TeV energy lepton-hadron colliders is mandatory to provide PDFs for adequate interpretation of forthcoming data from FCC-hh [6] and SppC [7]. Finally, energy-frontier lepton-hadron colliders are advantageous for investigation of a number of BSM phenomena (see, for example, Figure 1 for LHC, ILC and ILC*LHC comparison [8]).

Muon-proton colliders were proposed in 1990s (see reviews [9, 10] and references therein) as an alternative to linac-ring type electron-proton colliders (see reviews [11-14] and references therein). Muon-hadron colliders came back into the spotlight in the 2010s: the FCC based μp and μA colliders have been proposed in [15]; SppC based μp and μA colliders in [16] and [17], respectively; HL-LHC and HE-LHC based $μp$ and $μA$ colliders in [18] and [19], respectively; RHIC based $μp$ and μA colliders in [20]. Correct luminosity values keeping into account beam-beam tune-shifts and muon decays are presented in [10] for $μp$ colliders and in [19] for $μA$ colliders.

Almost all muon-hadron collider proposals assume construction of muon colliders (or dedicated muon ring) tangential to existing or proposed hadron colliders. In this respect, FNAL could provide an extraordinary opportunity because it has two tangential ring tunnels (see Fig. 1). The Tevatron ring can be used for proton beam and the Main injector for muon beam or vice-versa.

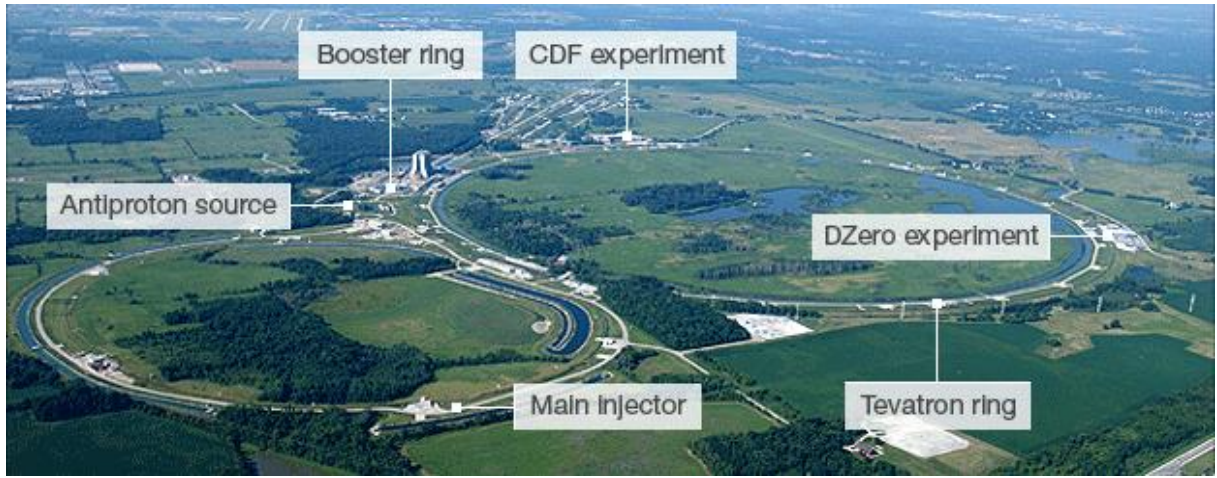

Figure 1. The FNAL Tevatron accelerator complex

In this letter we propose two possible options for muon-proton colliders at FNAL. In the next section, the main parameters of the first Tevatron-based μp collider, suggested a quarter of a century ago, are checked using AloHEP software [21, 22]. Then, in section 3 we discuss main parameters of the two proposed options for the Tevatron-based muon-proton collider. Finally, we present our conclusions and recommendations in Section 4.

**2. First proposal: 200 GeV muon ring tangential to Tevatron**

In 1996 Vladimir Shiltsev proposed construction of 200 GeV muon ring tangential to Tevatron 1000 GeV proton ring in order to investigate μp collisions at 894 GeV center-of-mass energy [23, 24]. Main parameters of proposed muon-proton collider are presented in Table 1.

Implementation of these parameters into the software AloHEP results in following luminosity values: $L=1.2\times10^{33}$ $cm^{-2}s^{-1}$ for high μ-rate and $L=1.0\times10^{32}$ $cm^{-2}s^{-1}$ for low μ-rate without considering losses resulting from muon decay. However, taking muon decays into account, AloHEP leads to $L=0.73\times10^{33}$ $cm^{-2}s^{-1}$ for high μ-rate and $L=0.6\times10^{32}$ $cm^{-2}s^{-1}$ for low μ-rate. Concerning beam-beam tune shift, AloHEP gives $\xi_\mu=0.027$ for both options, and $\xi_p=0.02$ for high rate and $\xi_p=0.0012$ for low rate. The last value is 2.4 times larger than the value given in Table 1, which is not so crucial.

Table 1. Two choices for muon beam parameters (Shiltsev's proposal)

| Parameter [Unit] | High µ-production | Low µ-production |
|---|---|---|
| Muon Energy [GeV] | 200 | 200 |
| Proton Energy [GeV] | 1000 | 1000 |
| # of muons per bunch [$10^{10}$] | 200 | 50 |
| Muon norm. emitt. [µm] | 50 | 200 |
| Storage turns | ~2000 | ~2000 |
| Muon pulse rate | 30 | 10 |
| Muons $\beta$ @ IP [cm] | 7.5 | 7.5 |
| Protons $\beta$ @ IP [cm] | 15 | 15 |
| Proton norm. emitt. [µm] | 12.5 | 50 |
| # of protons per bunch [$10^{10}$] | 125 | 500 |
| Proton tuneshift | 0.02 | 0.0005 |
| Muon tuneshift | 0.026 | 0.026 |
| Luminosity [$cm^{-2}s^{-1}$] | $1.3\times10^{33}$ | $1.1\times10^{32}$ |

## 3. Two options for *µp* collider at FNAL

Two possibilities for bending magnets are considered for both muon and proton rings, namely, 8 Tesla (LHC magnets) and 16 Tesla (FCC magnets).

### 3.1. First option: protons in Tevatron ring, muons in main injector

The beam energies that can be obtained by using 8 Tesla and 16 Tesla bending magnets are given in the Table 2 (let us remind that 8 Tesla bending magnets are used at the LHC and 16 Tesla magnets are proposed for FCC). Hereafter, we will refer to FNAL-based *µp* colliders as µ-FNAL.

Table 2. Beam energies and $\sqrt{s_{\mu p}}$ of the Tevatron based *µp* colliders

| Bending magnets | 8 Tesla | | 16 Tesla | |
|---|---|---|---|---|
| Beams | Proton | Muon | Proton | Muon |
| Circumference, km | 6.28 | 3.32 | 6.28 | 3.32 |
| Beam energy, TeV | 1.78 | 0.94 | 3.56 | 1.88 |
| Center of mass energy, TeV | 2.59 | | 5.17 | |

For comparison we present main parameters of the HERA, eRHIC and LHeC in Table 3. HERA was the first and unique ep collider operated at DESY between 1992 and 2007, eRHIC was recently approved by DoE and planned to operate in 2030s. It is seen that center-of-mass energy of μ-FNAL exceeds that of the LHeC by factors 2.2 and 4.4 for 8T and 16T magnets, respectively.

Table 3. Main parameters of the HERA, eRHIC and LHeC

| Collider | $E_l$, TeV | $E_p$, TeV | $\sqrt{s}$, TeV | L, cm$^{-2}$s$^{-1}$ |
|---|---|---|---|---|
| HERA | 0.0275 | 0.92 | 0.318 | $7.5\times10^{30}$ |
| e-RHIC | 0.01 | 0.275 | 0.105 | $1.05\times10^{34}$ |
| LHeC | 0.05 | 7 | 1.18 | $10^{34}$ |

Parameters of proton and muon beams used for estimation of luminosity values for Tevatron based *μp* colliders are presented in Table 4. Muon beam parameters for $E_\mu$ = 0.94 TeV and $E_\mu$ = 1.88 TeV correspond to $E_\mu$ = 0.75 TeV and $E_\mu$ = 1.5 TeV options of muon collider from [25], respectively; proton beam parameters correspond to HL-LHC option upgraded for ERL 60 GeV based *ep* collider (Table 2.11 in [6]). The number of proton bunches is determined by the synchronization of muon and proton bunches at the interaction point. The proton bunch spacing needs to be set at 40 meters to achieve synchronization during collisions in a proton ring that is 6.28 kilometers long and a muon ring that is 3.32 kilometers long. Which corresponds to 157 bunches for proton ring.

Table 4. Main parameters of proton and muon beams

| Bending magnets | 8 Tesla | | 16 Tesla | |
|---|---|---|---|---|
| Beams | Proton | Muon | Proton | Muon |
| Particle per bunch, $10^{11}$ | 2.2 | 20 | 2.2 | 20 |
| Bunches per beam | 157 | 1 | 157 | 1 |
| Normalized emittance $\varepsilon_N$, μm | 2.0 | 25 | 2.0 | 25 |
| β* at IP, cm | 7 | 1 | 7 | 0.5 |
| Repetition rate | - | 15 | - | 12 |

Implementing beam parameters from Table 4 into AloHEP software [21, 22], we obtain tune-shift and luminosity values presented in Table 5. It is seen that $\xi_p$ value is one order higher than acceptable value $\xi_p$ = 0.01. This value can be arranged with decreasing number of particles in muon bunch by a factor 12, which leads to corresponding decrease of luminosity. In principle, this can be compensated by increment of $N_p$. As a result, $\xi_\mu$ will increase by 12 times to a value of about 0.1, which is acceptable for a muon beam.

Let us mention that problems with muon generation and neutrino-induced radiation can be easier overcome comparing with corresponding muon colliders since the number of muons required for µ-FNAL is one order lower than for the muon collider. In addition, for muon-proton colliders one need only one beam instead of two opposite beams ($\mu^+$ and $\mu^-$) at muon colliders.

Table 5. Main parameters of the µ-FNAL's first option

| Bending magnets | 8 TESLA | 16 TESLA |
|---|---|---|
| $\sqrt{s}$, TeV | 2.57 | 5.17 |
| Tune shift $\xi_P$ | 0.12 | 0.12 |
| Tune shift $\xi_\mu$, $10^{-3}$ | 9.54 | 9.54 |
| Peak L, $10^{33}$ cm$^{-2}$s$^{-1}$ | 1.25 | 4.01 |
| L with decay, $10^{33}$ cm$^{-2}$s$^{-1}$ | 0.79 | 2.53 |

**3.2. Second option: muons in Tevatron ring, protons in main injector**

A decade ago, construction of a 3 TeV center-of mass energy muon collider at Tevatron ring have been proposed in [26] with a $10^{34}$ cm$^{-2}$s$^{-1}$ luminosity goal. In this case the main injector can be used for protons to handle muon-proton collisions. As a result, in Table 2, proton and muon beam energies will change places, while the muon-proton center-of-mass energies remain the same.

Parameters of proton and muon beams used for estimation of luminosity values for Tevatron based *µp* colliders are presented in Table 6. Muon beam parameters for $E_\mu$ = 1.78 TeV and $E_\mu$ = 3.56 TeV correspond to $E_\mu$ = 1.5 TeV and $E_\mu$ = 3.0 TeV options of muon collider from [25], respectively; proton beam parameters correspond to HL-LHC option upgraded for ERL 60 GeV based *ep* collider (Table 2.11 in [6]). Similar to the first option the number of proton bunches is determined by the synchronization of muon and proton bunches at the interaction point.

Table 6. Main parameters of proton and muon beams

| Bending magnets | 8 Tesla | | 16 Tesla | |
|---|---|---|---|---|
| Beams | Proton | Muon | Proton | Muon |
| Particle per bunch, $10^{11}$ | 2.2 | 20 | 2.2 | 20 |
| Bunches per beam | 83 | 1 | 83 | 1 |
| Normalized emittance $\varepsilon_N$, µm | 2.0 | 25 | 2.0 | 25 |
| $\beta^*$ at IP, cm | 7 | 0.5 | 7 | 0.25 |
| Repetition rate | - | 12 | - | 6 |

Implementing beam parameters from Table 6 into AloHEP software, we obtain tune-shift and luminosity values presented in Table 7. It is seen that while center-of-mass energies don't change, luminosity values are approximately half of that of the first option. Concerning tune-shift values the procedure described in previous subsection may be still applied.

Table 7. Main parameters of the μ-FNAL's second option

| Bending magnets | 8 TESLA | 16 TESLA |
|---|---|---|
| √s, TeV | 2.57 | 5.17 |
| Tune shift $\xi_P$ | 0.12 | 0.12 |
| Tune shift $\xi_\mu$, $10^{-3}$ | 9.54 | 9.54 |
| Peak L, $10^{32}$ cm$^{-2}$s$^{-1}$ | 5.30 | 10.6 |
| L with decay, $10^{32}$ cm$^{-2}$s$^{-1}$ | 3.35 | 6.70 |

If it is decided to construct the muon collider in the Tevatron ring, it seems reasonable to consider the possibility of establishing the muon-proton collider as the first stage for two reasons: (1) only one muon beam is needed and (2) the number of muons in the beam is an order lower than that for the muon collider.

## 4. Conclusion

Comparing Table 3 with Tables 5 and 7 it can be seen that in terms of center of mass energy, μ-FNAL is well above HERA and e-RHIC, and several times higher than LHeC. Certainly, detailed analysis of different physics phenomena should be performed in order to compare their physics search potentials. Evidently, μ-FNAL is unique for search for muon-related new physics: second family leptoquarks, vector-like leptons, contact interactions etc.

**Final note:** FNAL has sufficient human resources and infrastructure to build a TeV-scale high-luminosity muon-proton collider that will provide unique potential for both SM (especially the fundamental principles of QCD) and BSM searches.